\documentclass[twocolumn]{jpsj2} %% two-column layout
\usepackage[dvips]{color}

\title{
Quadrupolar Order and Structural Phase Transition in DyB$_4$ with Geometrical Frustration
}

\author{Daisuke \textsc{Okuyama}\thanks{oku@iiyo.phys.tohoku.ac.jp}, Takeshi \textsc{Matsumura}, Hironori \textsc{Nakao} and Youichi \textsc{Murakami}}

\inst{Department of Physics, Graduate School of Science, Tohoku University,
Sendai 980-8578, Japan}

\abst{
Structural phase transition accompanying with quadrupolar ordering in DyB$_4$
with Shastry-Sutherland type geometrical frustration has been studied by X-ray diffraction.
Previous study [D. Okuyama \textit{et al.}: J. Phys. Soc. Jpn. \textbf{74} (2005) 2434.] using resonant X-ray
scattering revealed short-range ordering of the $O_{zx}$-type quadrupolar moments and the $c$-plane
component of the magnetic moments in addition to long-range ordering of the $c$-axis component of the
magnetic moments.
The present report focuses on the lattice distortion below the quadrupolar ordering temperature
at $T_{\text{N2}}$=12.7 K.
The (0 0 $l$=integer) fundamental lattice reflection splits into four peaks along the $h$ and $k$ directions
and the ($h$=even 0 0) reflection becomes broad along the $l$ direction.
This indicates that a structural transition from tetragonal to monoclinic takes place below
$T_{\text{N2}}$ together with the ordering of the quadrupolar moments.
}

\kword{quadrupolar order, geometrical frustration, DyB$_4$}

\begin{document}
\maketitle

\section{Introduction} %% No sections necessary for express letters, letters and short notes

%%%---Introduction---%%%
Magnetic materials with geometrical frustration have attracted interest as systems with
various magnetic anomalies,
which arise from a situation where magnetic order is suppressed down to very low temperatures.
For example, CsCoCl$_3$ and CsNiCl$_3$ with a triangular lattice of Ising spins exhibit
an unusual antiferromagnetic state in which one of its three sublattices is disordered~\cite{Mekata77}.
Pyrochlore compounds, such as Dy$_2$Ti$_2$O$_7$ and Ho$_2$Ti$_2$O$_7$, do not
exhibit magnetic ordering down to 0.2 K and 0.35 K, respectively~\cite{Harris97,Bramwell01,Ramirez99}, 
because coexistence of a local uniaxial anisotropy and a ferro-type magnetic interaction causes
frustration and leads to spin liquid states.
%However, the frustrated state can easily be broken by external perturbations such as magnetic
%field and pressure, and long range magnetic order appears.~\citen{Harris97,Hertog00}
In the present report, we deal with a different type of geometrically frustrated system of DyB$_4$,
in which Dy ions form a Shastry-Sutherland lattice (SSL) with both magnetic and quadrupolar
degrees of freedom~\cite{Shastry81}.
%Shastry-Sutherland lattice, which form a triangle and square lattice, has been studied by
%theoretical approach;
%the phase diagrarms is made as the fuction of ration between the nearest neighbour and next nearest
%neighbour and the magnitude of the total angular momentum $J$~\cite{Shastry81}.
%SrCu$_2$(BO$_3$)$_2$ has an irregular triangle lattice and the quantum spin $S$=1/2 of Cu ions,
%resulting in the magnetic dimmer ground states~\cite{Kageyama99,Miyahara99},
%'±'ê'ç'ÌŒ‹‰Ê'́AShastry-Sutherland'̍l'¦'½phase diagram'ɏ€'¸'é'à'Ì'Å' 'éB
%In contrast, the compounds, DyB$_2$C and HoB$_2$C, with a regular triangle and square
%lattice and a large total angular momentum lead to a long ranged Neel order in the viewpoint
%of the phase diagram of Shastry and Sutherland~\cite{Shastry81}.
%However, the powder neutron scattering shows very broad peak at the low scattering angle.
%The Warren-type random layer theory was proposed the model, in which
%the magnetic random layers partly break the translational symmetry, resulting in
%the magnetic broad peaks at low scattering angle~\cite{Duijn03,Watanuki04,Watanuki05b}.
%This results do not support the phase diagram of Shastry and Sutherland.

\begin{figure}[t]
\begin{center}
\includegraphics*[width=8.7cm]{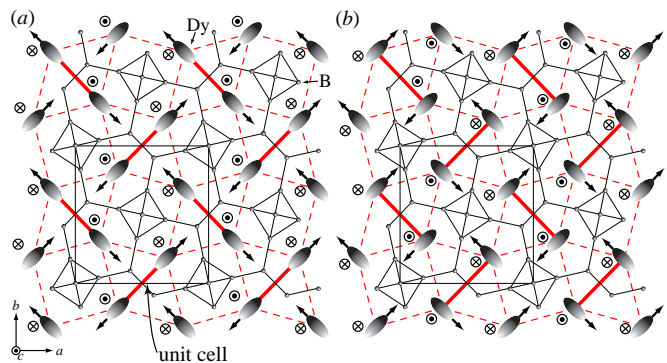}
\caption{
Models of possible magnetic and quadrupolar structures of DyB$_4$ in phase II and III proposed
from the results of neutron powder diffraction and resonant X-ray scattering~\cite{Watanuki05b,Okuyama05}.
Arrows represent $c$-axis and $c$-plane component of the magnetic moments.
Ellipses with light and dark shade represent $\langle O_{zx}\rangle$-type charge distributions
that are extended above and below the paper, respectively.
The local cartesian coordinates for each Dy ions are taken so that $x$ and $z$ axis corresponds
to the $c$-plane and $c$-axis magnetic component, respectively.
The nearest and next-nearest neighbor bonds between Dy ions are indicated by the thick lines
and dotted lines, respectively.
}
\label{fig1}
\end{center}
\end{figure}

%In DyB$_4$, Dy ions form a Shastry-Sutherland lattic (SSL), and it is expected that
%magnetic and quadrupolar interactions between Dy ions experience geometrical frustration.
DyB$_4$ has recently been attracting interest as a system where quadrupolar frustration
plays an important role.~\cite{Watanuki05a,Okuyama05}
The Dy ions are at the $4g$ sites of the tetragonal space group of $P4/mbm$:
$\mib{R}_{1}=(x, x+\frac{1}{2}, 0)$, $\mib{R}_{2}=(\frac{1}{2}-x, x, 0)$,
$\mib{R}_{3}=(-x, \frac{1}{2}-x, 0)$, and $\mib{R}_{4}=(\frac{1}{2}+x, -x, 0)$, with $x=0.3175$
for DyB$_4$.
The lattice constants are $a$=7.0960(4) \AA\ and $c$=4.0128(2) \AA\ at $T$=25 K~\cite{Watanuki05b}.
Basic physical properties of DyB$_4$ have been studied in detail by Watanuki \textit{et al}.
from the viewpoint of quadrupolar frustration~\cite{Watanuki05a,Watanuki05b}.
Two phase transitions take place at $T_{\text{N1}}$=20.3 K and $T_{\text{N2}}$=12.7 K.
Neutron powder diffraction and an analysis of magnetization show that the magnetic order
in phase II ($T_{\text{N2}}$ $\leq$ $T$ $\leq$ $T_{\text{N1}}$) is described by the propagation vector
$\mib{k}$=(1 0 0) with its moment along the $c$ axis.
It is suggested that a magnetic component within the $c$-plane appears in phase III ($T$ $\leq$ $T_{\text{N2}}$).
Specific heat measurement shows that the entropy of $R\ln{2}$ and $R\ln{4}$ is released
at $T_{\text{N2}}$ and $T_{\text{N1}}$, respectively, with increasing temperature.
This indicates that the ground state is a pseudo-quartet with two doublets closely located,
possessing both magnetic and quadrupolar degrees of freedom.
Elastic constant $C_{44}$, a strain susceptibility to a uniform distortion where $\angle ac$ or $\angle bc$
is modified from $90^{\circ}$, exhibits strong softening and ultrasonic attenuation takes place even in phase II,
where the magnetic moments along the $c$-axis are fixed.
This indicates that a quadrupolar degeneracy still remains and fluctuation of the quadrupolar moments
increases in phase II.
%Fitting of the elastic softening using a normal Curie-Weiss formula gives a positive Weiss temperature, which implies a ferro-type quadrupolar interaction.

%%%---the fundamental data of the compounds of subject matter---%%%
%%%These experimental results supports that the quadrupolar moments are strongly correlated but%%%
%%%strongly correlated due to the geometrical frustration.%%%
Magnetic and quadrupolar structures in phases II and III have been investigated in detail by resonant
X-ray scattering (RXS)~\cite{Okuyama05}.
Two possible model structures of these moments in phases II and III that are consistent with the experimental
results are illustrated in Fig. \ref{fig1}.
It is noted that these are single-$\mib{k}$ structures of (1 0 0), where the unit cell is the same as that 
of the chemical one.
The structures consist of a long-range order of the antiferromagnetic moment
along the $c$-axis ($\langle J_z \rangle$) and a short-range order of the magnetic component
within the $c$-plane ($\langle J_x \rangle$) and the quadrupolar moment $\langle O_{zx} \rangle$
that is compatible with $\langle \mib{J} \rangle$.
The short-range order observed by RXS and the elastic softening and absorption in phase II
suggests that the quadrupolar moments are strongly  correlated to exhibit short-range ordering
in a very short time scale of RXS measurement but fluctuating slowly in a time scale of elastic constant
measurement. This we consider to be due to geometrical frustration.

%%%---issues related to subject---%%%
In the present paper, we report on the structural phase transition in phase III that accompany
with the $O_{zx}$-type quadrupolar order.
Peak profiles of the (1 0 0) forbidden reflection at resonance that observes
the orderings of magnetic and quadrupolar moments are also reported.

\section{Experiment}

%%%---experimental method---%%%
%%%About the samples and experimental conditions%%%
A single crystal was grown by the floating zone method using a high-frequency furnace.
The quality of the sample was checked by magnetic susceptibility, which showed good agreement
with the result in ref.~\citen{Watanuki05a}.
Experiments were performed using an X-ray beam from a rotating anode of Mo target (17.48 keV)
at Tohoku University to study lattice distortion
and using a synchrotron X-ray at BL-16A2 of the Photon Factory in KEK to study orderings of
magnetic and quadrupolar moments.
The experiments were performed on a four-circle diffractometer using a sample with polished (0 0 1)
and (1 0 0) surfaces.

\begin{figure}[t]
\begin{center}
\includegraphics*[width=8cm]{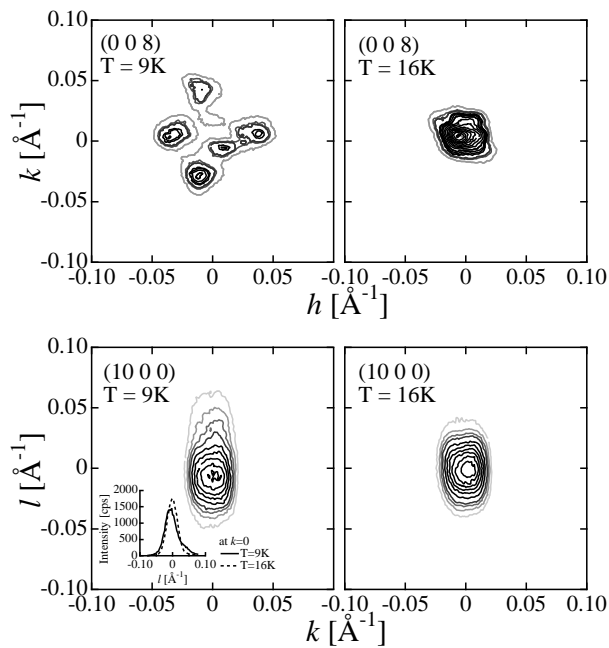}
\caption{
Contour maps of the intensity of the (0 0 8) and (10 0 0) reflections measured in the nonresonant region 
at E=17.48 keV. 
}
\label{fig2}
\end{center}
\end{figure}

%%%---experimental result---%%%
%%%Contour maps of the fundamental reflection show that%%%
%%%the phase transition is from tetragonal to monoclinic%%%
Figure \ref{fig2} shows contour maps of the intensity of the fundamental Bragg reflections of (0 0 8)
and (10 0 0).
The single peak of (0 0 8) in phase I and II splits into four peaks along the $h$ and $k$ directions in phase III.
The (10 0 0) peak becomes slightly broad along the $l$ direction.
These experimental results and a survey of some other Bragg reflections, (0 0 4), (0 0 6), and (0 0 10),
show that the structural phase transition is from tetragonal to monoclinic.
The angle $\angle ac$, or $\angle bc$, is estimated to be 89.84$^{\circ}$ at 9 K.
The spliting of the peak is due to the formation of domains of the monoclinic distortion.
We consider that the peak in (0 0 8) that remains at around the center arose from another domain
in which the $a$-axis did not move but the $c$-axis moved,
which is reflected in the broadening of the (10 0 0) peak along the $l$ direction.

\begin{figure}[t]
\begin{center}
\includegraphics*[width=8cm]{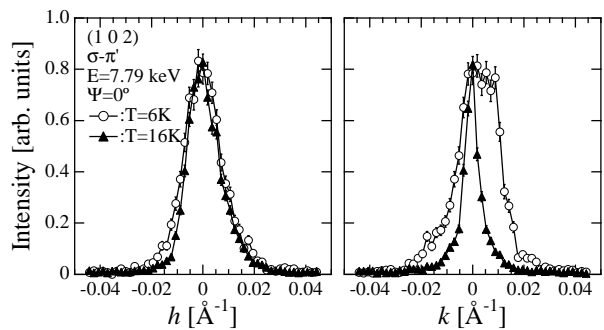}
\caption{
Peak profile of the (1 0 2) resonant magnetic reflection measured at the $L_{\mathrm{III}}$
absorption edge of Dy at $\Psi=0^{\circ}$.
}
\label{fig3}
\end{center}
\end{figure}

%%%---experimental result---%%%
%%%Magnetic propagation vectors result in the two domain of the monoclinic distortion%%%
It should be remarked that the antiferromagnetic domains with the (1 0 0) and (0 1 0) propagation vectors
result in the monoclinic distortions within the $bc$ plane ($\angle bc \neq 90^{\circ}$) and 
$ac$ plane ($\angle ac \neq 90^{\circ}$), respectively.
This is inferred from the fact that the (1 0 2) resonant magnetic Bragg peak splits only
along the $k$ direction in phase III as shown in Fig. \ref{fig3}.
The peak profile along the $h$ direction exhibits no difference between 16 K and 6 K.

\begin{figure}[t]
\begin{center}
\includegraphics*[width=8cm]{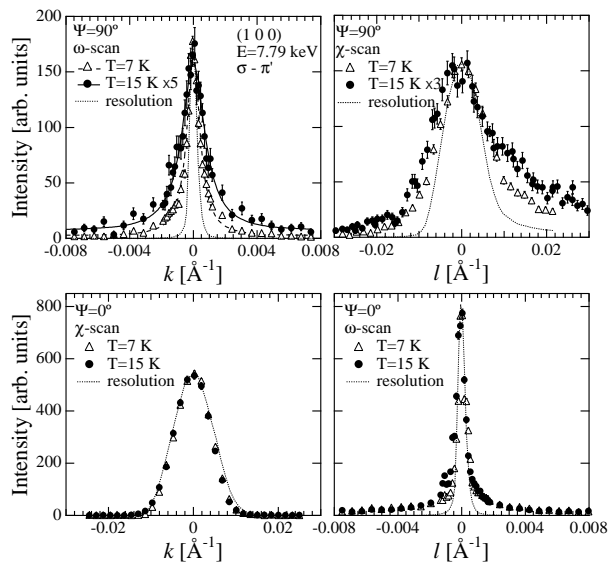}
\caption{
Resonant peak profiles of the (1 0 0) forbidden reflection at $\Psi$=0$^{\circ}$ and 90$^{\circ}$
measured at the $L_{\mathrm{III}}$ absorption edge of Dy.
T=7 K and 15K correspond to phase III and II, respectively.
The resolution functions represented by the thin dotted lines were estimated from the (2 0 0) fundamental
Bragg reflection at T=15 K.
Note the difference in the resolution width for $\omega$-scans and $\chi$-scans.
}
\label{fig4}
\end{center}
\end{figure}

%%%---experimental result---%%%
%%%Contour maps of the fundamental reflection show that%%%
Figure \ref{fig4} shows the temperature dependences of the profiles of the (1 0 0) resonant reflection
at azimuthal angles of $\Psi$=0$^{\circ}$ and 90$^{\circ}$.
The azimuthal angle is defined to be zero when the $c$-axis is within the scattering plane.
Obviously, the peak widths at $\Psi$=90$^{\circ}$ are broader than the resolution both for
the $k$ and $l$ directions.
In contrast, the widths at $\Psi$=0$^{\circ}$ are limited by the resolution.
An analysis of the resonant reflection in ref. \citen{Okuyama05} shows that the resonance
at $\Psi$=0$^{\circ}$ arises from the antiferromagnetic ordering of $\langle J_z \rangle$ 
and at $\Psi$=90$^{\circ}$ from the ordering of $\langle J_x \rangle$ and $\langle O_{zx} \rangle$,
because the structure factor of the resonant diffraction is proportional to $\langle J_z \rangle \cos{\Psi}$,
$\langle J_x \rangle \sin{\Psi}$, and $\langle O_{zx} \rangle \sin{\Psi}$.
%In addition, the correlation lengths of each order parameter were discussed in detail.
%The broad and sharp profiles of (1 0 0) reflection are fitted by a Lorentzian function, which is convoluted 
%with the resolution function.
%The broad profiles at $\Psi$=90$^{\circ}$ along $b$-axis, corresponding to short-range ordering of
%$\langle J_x \rangle$ and $\langle O_{zx} \rangle$, are estimated to be 
%(1.4$\pm$0.1)$\times$10$^{3} \AA$ at 15 K in phase II and (4.2$\pm$0.2)$\times$10$^{3} \AA$ 
%at 7 K in phase III.
%These fitting llines at 15 K and 7 K are described by a solid and thick dotted line, respectively.
%The profiles at $\Psi$=90$^{\circ}$ along $c$-axis are also broader than the resolution function,
%although precise estimation was difficult.
%On the other hand, the sharp profiles at $\Psi$=0$^{\circ}$, corresponding to long-range ordering of 
%$\langle J_z \rangle$, are almost as sharp as that of the resolution limit, 2$\times$10$^{4} \AA$.
%The profiles at $\Psi$=90$^{\circ}$ along the $c$-axis are also broader than the resolution function,
%although precise estimation was difficult.
The correlation length of $\langle J_x \rangle$ and $\langle O_{zx} \rangle$ along the $b$-axis
estimated from the fitting of the peak profile along the $k$ direction at $\Psi$=90$^{\circ}$
is (1.4$\pm$0.1)$\times$10$^{3}$ \AA\ at 15 K in phase II and (4.2$\pm$0.2)$\times$10$^{3}$ \AA\ 
at 7 K in phase III~\cite{Okuyama05}.
The solid and dashed lines in Fig. \ref{fig4} represent the fitting lines convoluted with the resolution.
In contrast, the correlation length of $\langle J_z \rangle$ along the $c$-axis
%estimated from the fitting of the peak profile along the $l$ direction 
at $\Psi=0^{\circ}$ is $(9.2\pm0.5)\times 10^3$ \AA\ at 15 K in phase II and over $2\times 10^4$ \AA, 
the resolution limit, at 7 K in phase III~\cite{Okuyama05}.
It should be noted that the peak profile of the (1 0 0) reflection in phase III is little affected by the
monoclinic distortion because the magnetic domain with the (1 0 0) propagation vector
is distorted within the $bc$-plane.
We consider that the broadened peak width intrinsically demonstrates the short-ranged correlation length
of the order parameter.
%The $k$-scan of (1 0 0) reflection at $\Psi$=90$^{\circ}$ corresponding to
%$\omega$-scan is clearly shown that the broad peak at T=15K in phase II becomes
%sharp at 7K in phase III, which do not diverge over the resolution limit.
%These experimental results support strongly ref. \citen{Okuyama05}.

\section{Discussion}

Let us first discuss the mechanism of the lattice distortion. 
One possible scenario is that the frustration is released by modifying the SSL of Dy ions 
as a result of the lattice distortion. 
However, the uniform monoclinic distortion in phase III little modifies the SSL configuration 
within the $c$ plane.
Furthermore, the peak width of the (1 0 0) reflection at $\Psi$=90$^{\circ}$ do not diverge 
to the resolution limit after the structural phase transition,
which implies that the effect of frustration is still alive.
For these reasons, this scenario is not likely to be the case. 
Another one could be that the ordering of $\langle O_{zx} \rangle$ moments induces the 
distortion through cooperative Jahn-Teller effect.
The arrangements of $\langle O_{zx} \rangle$ as illustrated in Fig. \ref{fig1} will cooperatively
favor a monoclinic distortion where $\angle bc$ is modified.
This is consistent with the experimental results of X-ray diffraction in phase III and 
also consistent with the huge elastic softening of the $C_{44}$ mode in phase II. 
Then, we consider that the monoclinic distortion is caused by the ordering of the 
$\langle O_{zx} \rangle$ moments. 

An important problem in DyB$_4$ is why the doublet remains even below $T_{\text{N1}}$ 
where the magnetic moment along the $c$-axis is fixed. The appearance of the ordered magnetic 
moment normally splits the quasi-quartet into four singlets via the Zeeman effect. 
However, a doublet somehow remains and both quadrupolar and magnetic degrees 
of freedom survive within the $c$-plane. These moments are short-range correlated in phase II 
as observed in RXS, but it is not static because there is elastic softening and absorption. 
We suggest that they are fluctuating in the time scale of ultrasonic measurement. 
Below $T_{\text{N2}}$ the correlation becomes static and the elastic softening stops. 

Another point of importance is that the correlation length do not diverge as can be observed 
in the peak profile in Fig. \ref{fig4} at 7 K and $\Psi$=90$^{\circ}$. 
However, the correlation length of the order of 10$^3$ \AA\ is quite long to mention it as short-ranged. 
We consider that the broadened profile itself might be one of the characteristics of SSL. 
The sharp specific heat anomaly and the entropy release of $R\ln{2}$ may be due to this 
correlation of the order of 10$^3$ \AA. 
There might be a novel mechanism that prevent complete long-range order in the SSL of rare-earth
tetraborides.  
%However, the correlation length do not diverge as can be observed in the peak width in Fig. \ref{fig4} 
%at 7 K and $\Psi=90^{\circ}$. 
%We consider that the correlation length 4.2$\times$10$^{3} \AA$ at 7 K in phase III is long enough to 
%release $R\ln{2}$ entropy.
%The broadened profile itself might be one of the characteristic of SSL.
%
%In spite of the geometrically frustrated structure of DyB$_4$, 
%the ordering temperatures of $T_{\text{N1}}$=20.3 K and $T_{\text{N2}}$=12.7 K are quite 
%normal in comparison with other Dy compounds.
%This point is different from triangular Ising lattices and pyrochlore compounds we referred to in the 
%introduction. What is intriguing in DyB$_4$ is that the correlation length of the in-plane moment 
%$\langle J_x \rangle$ and $\langle O_{zx} \rangle$ do not diverge even in phase III where the order 
%is expected to be static. 
%There might be a novel mechanism that prevent long-range order in the SSL of rare earth tetraborides. 

\section{Summary}

We performed resonant and nonresonant X-ray diffraction experiments in DyB$_4$,
in which the rare earth ions form a Shastry-Sutherland lattice.
Below $T_{\text{N2}}$=12.7 K, the (0 0 8) fundamental lattice reflection splits into four 
peaks along the $h$ and $k$ directions in the reciprocal space and
the (10 0 0) reflection becomes slightly broader along the $l$ direction. 
These results indicate that the structural transition is from tetragonal to monoclinic, 
which is consistent with the model structure of the ordering of $O_{zx}$-type quadrupolar moment. 
The correlation length of the magnetic and quadrupolar moments in phase II, which is 
estimated from the peak width of the (1 0 0) resonant reflection at $\Psi$=90$^{\circ}$, 
is obviously shorter than the resolution limit, and do not diverge even in phase III where 
the order becomes static.

\section*{Acknowledgment}
The authors are indebted to R. Watanuki for many fruitful discussions.
We also thank S. Kunii, K. Horiuchi, M. Onodera, H. Shida for assistance in single-crystal growth,
and K. Iwasa and K. Suzuki for profitable discussions.
This study was performed with the approval of the Photon Factory Program Advisory
Committee (No. 2004G235), and was supported by the 21st century center of excellence program, 
and by a Grant-in-Aid for Scientific Research from the Japanese Society for the Promotion of Science.


\begin{thebibliography}{99} %% The number "99" means that this list has more than nine items.
\bibitem{Mekata77} M. Mekata: J. Phys. Soc. Jpn. \textbf{42} (1977) 76.
\bibitem{Harris97} M. J. Harris, S. T. Bramwell, D. F. McMorrow, T. Zeiske and K. W. Godfrey:
Phys. Rev. Lett. \textbf{79} (1997) 2554.
\bibitem{Bramwell01} S. T. Bramwell, M. J. Harris, B. C. den Hertog, M. J. P. Gingras, J. S. Gardner,
D. F. McMorrow, A. R. Wildes, A. L. Cornelius, J. D. M. Champion, R. G. Melko and T. Fennell:
Phys. Rev. Lett. \textbf{87} (2001) 047205.
\bibitem{Ramirez99} A. P. Ramirez, A. Hayashi, R. J. Cava, R. Siddharthan and B. S. Shastry:
Nature \textbf{399} (1999) 333.
%\bibitem{Hertog00} Byron. C. den Hertog and Michel J. P. Gingras:
%Phys. Rev. Lett. \textbf{84} (2000) 3430.
%\bibitem{Kageyama99} H. Kageyama, K. Yoshimura, R. Stern, N. V. Mushinikov, K. Onizuka, M. Kato,
%K. Kosuge, C. P. Slichter, T. Goto and Y. Ueda: Phys. Rev. Lett. \textbf{82} (1999) 3168.
%\bibitem{Miyahara99} S. Miyahara and K. Ueda: Phys. Rev. Lett. \textbf{82} (1999) 3701.
%\bibitem{Duijn03} J. van Duijn, J. P. Attfield, R. Watanuki, K. Suzuki and R. K. Heenan:
%Phys. Rev. Lett. \textbf{90} (2003) 087201.
%\bibitem{Watanuki04} R. Watanuki, K. Suzuki, J. van Duijn and J. P. Attfield:
%Phys. Rev. B \textbf{69} (2004) 064433.
\bibitem{Shastry81} B. S. Shastry and B. Sutherland: Physica \textbf{108}B (1981) 1069.
\bibitem{Watanuki05a} R. Watanuki, G. Sato, K. Suzuki, M. Ishihara, T. Yanagisawa, Y. Nemoto
and T. Goto: J. Phys. Soc. Jpn. \textbf{74} (2005) 2169.
\bibitem{Okuyama05} D. Okuyama, T. Matsumura, H. Nakao and Y. Murakami: J. Phys. Soc. Jpn. \textbf{74} (2005) 2434.
\bibitem{Watanuki05b} R. Watanuki: Ph.D. thesis, Yokohama National University (2004).
%\bibitem{MyExperiment05_1} We call the magnetic components parallel and perpendicular to
%the $c$-axis as the $c$-axis component and $c$-plane component, respectively.
\end{thebibliography}
\end{document}